\documentclass[aps,prl,twocolumn,groupedaddress,showpacs]{revtex4} 

\usepackage{graphicx}
\usepackage{dcolumn}
\usepackage{bm}

\begin{document} 

\title{Electronic Inhomogeneity and Breakdown of the 
Universal Thermal Conductivity in Cuprate Superconductors}

\author{X. F. Sun} 
\author{S. Ono} 
\author{Yasushi Abe} 
\altaffiliation{Present address: CERC,AIST, Tsukuba 205-8562, 
Japan} 
\author{Seiki Komiya} 
\author{Kouji Segawa} 
\author{Yoichi Ando} 
\affiliation{Central Research Institute of Electric Power 
Industry, Komae, Tokyo 201-8511, Japan} 

\date{\today} 

\begin{abstract} 

We report systematic, high-precision measurements of the low-$T$ (down
to 70 mK) thermal conductivity $\kappa$ of YBa$_2$Cu$_3$O$_y$,
La$_{2-x}$Sr$_x$CuO$_4$ and Bi$_2$Sr$_2$CaCu$_2$O$_{8+\delta}$. Careful
examinations of the Zn- and hole-doping dependences of the residual
thermal conductivity $\kappa_0/T$, as well as the in-plane anisotropy of
$\kappa_0/T$ in Bi$_2$Sr$_2$CaCu$_2$O$_{8+\delta}$, indicate a breakdown
of the universal thermal conductivity, a notable theoretical prediction
for $d$-wave superconductors. Our results point to an important role of
electronic inhomogeneities, that are not considered in the standard
perturbation theory for thermal conductivity, in the under- to
optimally-doped regime.

\end{abstract} 

\pacs{74.25.Fy, 74.25.Dw, 74.72.Bk, 74.72.bk, 74.72.Dn} 

\maketitle 

The ``gapless" nature of the $d$-wave superconductivity in high-$T_c$
cuprates opened a new possibility to study the electronic state at low
temperature through quasiparticle (QP) heat transport \cite{Taillefer,
Chiao1, Chiao2, Nakamae1, Hussey, Takeya, Sutherland, Hawthorn_Tl, Hill,
Ando1, Sun1, Sun2, Hawthorn}. In the presence of disorder, QPs are
created near the gap nodes with an ``impurity bandwidth" $\gamma$
\cite{Hussey_Review} and carry heat; intriguingly, those QPs are also
scattered by the same disorder, and in some cases the effects of
creation and scattering balance, causing the thermal conductivity
$\kappa$ due to QPs to become independent of the disorder concentration
at low $T$. This phenomenon is called the ``universal thermal
conductivity", which is best understood in the framework of the
self-consistent $T$-matrix approximation (SCTMA) theory
\cite{Hussey_Review} for BCS $d$-wave superconductors. In the SCTMA
theory, when $\gamma$ is in the universal limit $k_BT \ll \gamma \ll
\Delta_0$ ($\Delta_0$ is the maximum gap), the residual component
$\kappa_0/T$ ($T \rightarrow$ 0 limit of $\kappa/T$) becomes independent
of the QP scattering rate $\Gamma$ and is expressed as \cite{Graf,
Durst}
\begin{equation}
\frac{\kappa_0}{T} = \frac{k_B^2}{3 \hbar} \frac{n}{c} \left( 
\frac{v_F}{v_2} + \frac{v_2}{v_F} \right) \simeq
\frac{k_B^2}{3 \hbar} \frac{n}{c} \frac{v_F}{v_2}, 
\label{k0/T}
\end{equation}
with $n$ the number of $\rm{CuO_2}$ planes per unit cell, $c$ the 
$c$-axis lattice constant, and $v_F$ ($v_2$) the QP velocity 
normal (tangential) to the Fermi surface at the node. This is 
useful because, if valid, one can obtain $\Delta_0$ 
by the bulk measurement of $\kappa$ \cite{Sutherland,Hawthorn_Tl}.

Experimentally, $\kappa_0/T$ was reported to be approximately
independent of the impurity scattering in nearly optimally-doped
YBa$_2$Cu$_3$O$_y$ (YBCO) \cite{Taillefer} and
Bi$_2$Sr$_2$CaCu$_2$O$_{8+\delta}$ (Bi2212) \cite{Nakamae1}, and its
magnitude for Bi2212 was consistent \cite{Chiao2} with Eq. (\ref{k0/T})
using the angle-resolved photoemission spectroscopy (ARPES) results
\cite{ARPES}. Hence, it has been considered that the universal thermal
conductivity is essentially realized in cuprates. However, the
experimental results so far \cite{Taillefer, Chiao1, Chiao2, Nakamae1}
carry rather large error bars and thus it has not been clear to what
extent $\kappa_0/T$ is ``universal". Moreover, Hussey {\it et al.}
\cite{Hussey} found an {\it absence} of $\kappa_0/T$ in an underdoped
YBa$_2$Cu$_4$O$_8$, which questions the universal scenario and suggests
a QP localization phenomenon that does not show up in the SCTMA theory.
In this regard, an unusual QP localization in magnetic field was
recently observed in underdoped La$_{2-x}$Sr$_x$CuO$_4$ (LSCO)
\cite{Sun2, Hawthorn} and, while it is obviously related to the
magnetic-field-induced spin/charge order \cite{Sun2, kappa_H}, the
mechanism of this localization is not yet understood. 

Theoretically, whereas the SCTMA theory is a perturbation theory, 
there are other non-perturbative approaches \cite{Hussey_Review, 
Atkinson, Senthil} that tend to predict QP localizations. 
Actually, when the superconducting (SC) state is inhomogeneous, 
the SCTMA theory is clearly inappropriate and nonperturbative 
theories becomes necessary. Since the cuprate superconductors 
have an inherent tendency to develop electronic inhomogeneities 
in the SC state because of the short coherence length, and indeed 
various sorts of electronic inhomogeneities are being discovered 
in cuprates \cite{McElroy, Pan, Vershinin, Howald}, it is 
important to sort out the consequence of the inhomogeneity in the 
low-energy physics probed by the QP heat transport.

In this Letter, we critically examine the universal thermal conductivity
by studying YBCO, LSCO, and Bi2212 for various doping levels; what is
crucial to this work is that our samples are among the best single
crystals available for these three systems, well characterized by
various transport properties \cite{Segawa, Komiya, Komiya_Zn, Ando2,
Ando3, Basov}, and the $\kappa$ measurements are done with a small
absolute uncertainty of less than 10\%. In our experiments, we find
three features that all indicate a breakdown of the universal thermal
conductivity upon a closer look: i) a slight Zn substitution for Cu
induces notable suppression of $\kappa_0/T$ in underdoped and
optimally-doped samples of all the three systems; ii) in Bi2212, both the
magnitude and the doping dependence of the gap parameter $v_F/v_2$
obtained from $\kappa_0/T$ via Eq. (\ref{k0/T}) differ significantly
from the ARPES results \cite{ARPES, Mesot}; iii) a large in-plane
anisotropy of $\sim$2 is observed for $\kappa_0/T$ in Bi2212. We discuss
that the electronic inhomogeneity plays a major role in the departure
from the universality by causing a partial QP localization as suggested
by the nonperturbative theories. 

\begin{figure}
\includegraphics[clip,width=8.0cm]{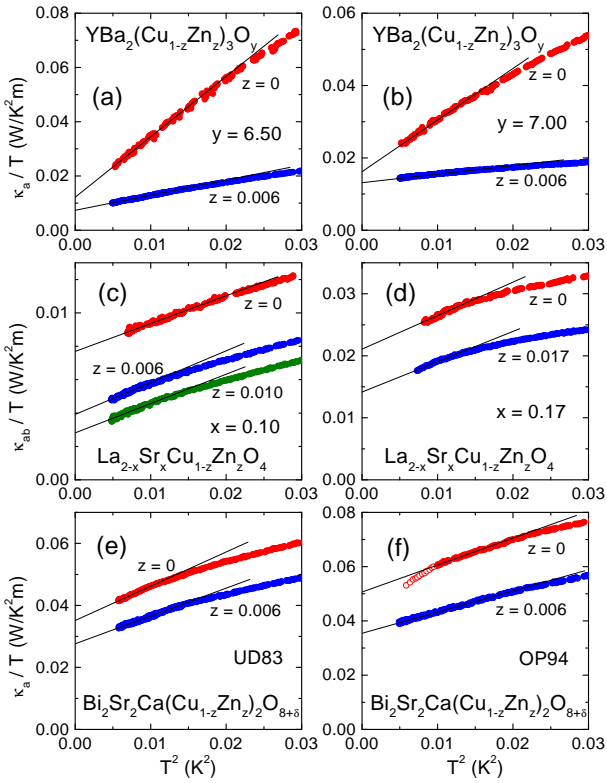} 
\caption{(color online) Zn-substitution effect on the low-$T$ 
thermal conductivity of YBCO, LSCO, and Bi2212 in the underdoped 
and optimally-doped regimes. 
The doping levels of Bi2212 samples are indicated by the $T_c$ 
values of pure ones (Zn-substituted ones are annealed to achieve 
the same oxygen contents as the pure ones). 
The solid lines are linear fits to the lowest-$T$ data to extract 
$\kappa_0/T$, except for the $z$ = 0 sample in (f) (see text).} 
\end{figure} 

High-quality single crystals of YBa$_2$(Cu$_{1-z}$Zn$_z$)$_3$O$_y$,
La$_{2-x}$Sr$_x$Cu$_{1-z}$Zn$_z$O$_4$ and
Bi$_2$Sr$_2$Ca(Cu$_{1-z}$Zn$_z$)$_2$O$_{8+\delta}$ are grown by a flux
method \cite{Segawa}, a traveling-solvent floating-zone method
\cite{Komiya} and a floating-zone method \cite{Ando2, Ando3},
respectively. The actual Zn concentration $z$ is determined by the
inductively-coupled plasma atomic-emission spectroscopy (ICP-AES). All
the YBCO samples are perfectly detwinned and $\kappa$ is measured along
the $a$ axis to avoid the additional electronic heat transport coming
from the Cu-O chains. For a doping-dependence study of Bi2212,
Bi$_2$Sr$_2$CaCu$_2$O$_{8+\delta}$ crystals as well as
Bi$_2$Sr$_2$Ca$_{1-y}$Dy$_y$Cu$_2$O$_{8+\delta}$ (Dy-Bi2212) crystals
(also produced by the floating-zone method) are carefully annealed at
400--800$^{\circ}$C in suitable atmospheres to tune the oxygen
content. We label the Bi2212 samples as UD70, OP94, OD70, {\it etc.},
by their doping regimes and $T_c$ values, and the Dy-Bi2212 samples as
Dy81, Dy45, {\it etc.}, by their $T_c$ values (all in underdoped regime)
\cite{note_Dy}. The thermal conductivity measurement in millikelvin
region is done by a conventional steady-state ``one heater, two
thermometer" technique in a dilution refrigerator \cite{Takeya, Sun1}.
We emphasize that our $\kappa$ data are reproducible with typical
variations of less than 10\%, thanks to the consistently high quality
of the crystals and a good control of the doping level
\cite{Segawa, Komiya, Komiya_Zn, Ando2, Ando3}. 

First, we show the effect of Zn-substitution on the low-$T$ QP 
heat transport for different doping regimes. Figure 1 shows the 
$\kappa/T$ vs $T^2$ plots for YBCO, LSCO, and Bi2212 systems in 
the underdoped and optimally-doped regimes. The $T$=0 intercepts 
of the linear fits to the lowest-$T$ data give the residual QP 
component $\kappa_0/T$ \cite{Taillefer, Chiao1, Chiao2, Nakamae1, 
Hussey, Takeya, Ando1, Sun1}. Here, the striking feature is a 
strong {\it decrease} of $\kappa_0/T$ upon slight Zn substitution 
in all these systems \cite{note_slope}, whereas within the SCTMA 
theory $\kappa_0/T$ is predicted to {\it increase} with $\Gamma$ 
when the system is not in the universal limit \cite{Graf, Maki}. 
We note that in YBCO at $y$ = 7.00, $\kappa_0/T$ decreases by 
$\sim$20\% with 0.6\% of Zn [Fig. 1(b)]; while this behavior 
suggests a departure from the universal behavior given our small 
error bar ($\sim$10\%) \cite{note_error}, it is not inconsistent 
with the previous results by Taillefer {\it et al.} for $y$ = 6.9 
\cite{Taillefer, Chiao1}, where the error bar was as large as 
$\sim$40\%. A more pronounced suppression in $\kappa_0/T$ of 
$\sim$40\% is observed upon 0.6\% of Zn-substitution in 
underdoped YBCO at $y$ = 6.50 [Fig. 1(a)], for which there has 
been no previous report. Apparently, the present higher-precision 
measurements indicate that all three systems do not strictly 
display the universal behavior in the underdoped and 
optimally-doped regimes; furthermore, the observed {\it 
suppression} of $\kappa_0/T$ by Zn substitution is opposite to 
what is expected from the SCTMA theory. If we remember that a 
non-perturbative theory that takes into account the gap 
inhomogeneities induced by impurities predicts a QP localization 
\cite{Atkinson}, it seems most likely that the observed behavior 
is essentially due to the Zn-induced electronic inhomogeneity in 
the form of nonsuperconducting droplets \cite{Pan}.

\begin{figure}
\includegraphics[clip,width=8.0cm]{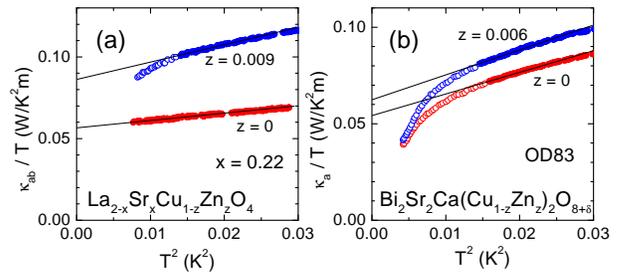} 
\caption{(color online) 
Zn-substitution effect on the low-$T$ thermal conductivity of LSCO 
and Bi2212 in the overdoped region. The data that are apparently 
affected by the electron-phonon decoupling (see text) are shown by 
open symbols. 
The solid lines are linear fits to estimate $\kappa_0/T$.}
\end{figure} 

Figure 2 shows the Zn-substitution effect on $\kappa$ in overdoped LSCO
and Bi2212. In overdoped cuprates, serious electron-phonon decoupling
occurs at very low $T$ \cite{Smith, Nakamae2}, causing the QP thermal
conductivity to be undetectable by the conventional steady-state
technique. As a result, a strong downturn shows up in the lowest-$T$
data of $\kappa/T$ for overdoped samples \cite{Smith, Nakamae2}, which
is also the case with our data in Fig. 2. [Our optimally-doped Bi2212
also shows this problem below 100 mK, see Fig. 1(f)]. It is therefore
impossible to precisely determine $\kappa_0/T$ of these samples from the
present data. Nevertheless, for a qualitative evaluation of the
Zn-substitution effect, one can crudely estimate $\kappa_0/T$ by a
linear fitting to a higher temperature range, as shown in Fig. 2, and it
would be safe to conclude that upon Zn substitution $\kappa_0/T$ is
{\it enhanced} in these overdoped samples. This trend is opposite to
that observed in underdoped and optimally-doped samples in Fig 1, and is
probably understandable within the SCTMA theory \cite{Graf, Maki}.
Hence, there is a crossover near optimum doping beyond which the nodal
QPs behave more ordinarily.

\begin{figure}
\includegraphics[clip,width=7.5cm]{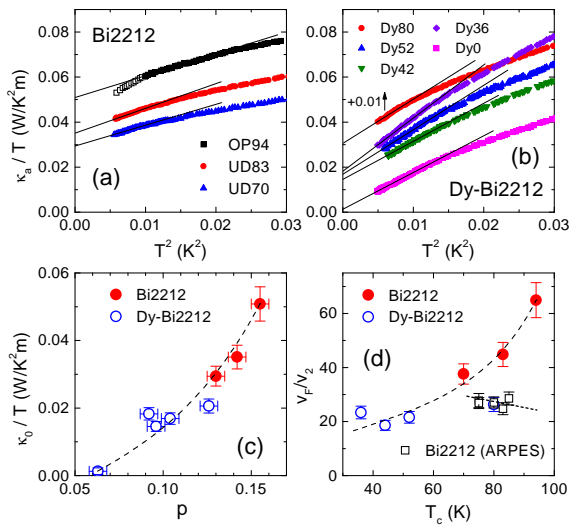} 
\caption{(color online) (a),(b) $a$-axis thermal conductivity of 
Bi2212 and Dy-Bi2212 at low $T$; Dy80 data are shifted up by 0.01 
W/K$^2$m for clarity. The solid lines are linear fits to extract 
$\kappa_0/T$. (c) Hole-doping dependence of $\kappa_0/T$ in Bi2212. 
(d) The gap parameter $v_F/v_2$, calculated from 
$\kappa_0/T$ using Eq. (\ref{k0/T}), as a function of $T_c$ in the 
underdoped region; also shown is $v_F/v_2$ obtained from ARPES 
\cite{Mesot} for comparison. The dashed and dotted lines are 
guides to the eyes.}
\end{figure} 

We next show the doping dependence of $\kappa_0/T$ for Bi2212, 
where the gap parameters have been reliably obtained by ARPES 
studies \cite{ARPES, Mesot}. Figures 3(a) and 3(b) show the 
low-$T$ thermal conductivity (along the $a$ axis) of Bi2212 and 
Dy-Bi2212 single crystals, and Fig. 3(c) summarizes the doping 
dependence of $\kappa_0/T$. Here, the hole concentration per Cu, 
$p$, is determined by the method employing the Hall coefficient 
proposed in Ref. \cite{Ando4}. Clearly, $\kappa_0/T$ decreases 
with decreasing doping and approaches zero upon entering the 
nonsuperconducting region --- this behavior is essentially the 
same as that observed in LSCO \cite{Takeya, Sutherland} and YBCO 
\cite{Sun1}. The gap parameter $v_F/v_2$ calculated using Eq. 
(\ref{k0/T}) is shown in Fig. 3(d); also included in this panel 
are the $v_F/v_2$ values directly obtained from ARPES for several 
underdoped Bi2212 crystals \cite{Mesot}. Not only the opposite 
trend in the doping dependence, but also the difference in the 
$v_F/v_2$ values from $\kappa_0/T$ and from ARPES, demonstrate an 
inadequacy of Eq. (\ref{k0/T}). To be more precise, $\kappa_0/T$ 
of 0.051 W/K$^2$m for the optimally-doped Bi2212 yields $v_F/v_2$ 
= 65 and $\Delta_0$ = 10 meV (assuming a simple $d_{x^2-y^2}$ 
gap), in contrast to $v_F/v_2$ = 20 and $\Delta_0$ = 35 meV 
obtained from ARPES \cite{ARPES, Mesot}. The reason why 
$\kappa_0/T$ is much larger than what the ARPES-derived $v_F/v_2$ 
would predict via Eq. (\ref{k0/T}) is not clear, but this should 
be understood in relation with the electronic inhomogeneity, 
given that there is likely a large variation in the SC gap 
magnitude in this material \cite{McElroy, Howald}. It might be 
that, depending on the nature of the inhomogeneity, sometimes the 
QP creation effect overwhelms the QP scattering effect (which may 
be the case near optimum doping) and sometimes the opposite 
happens (possibly in the underdoped regime). Clearly, more work 
is called for on the theoretical part.

\begin{figure} 
\includegraphics[clip,width=7.5cm]{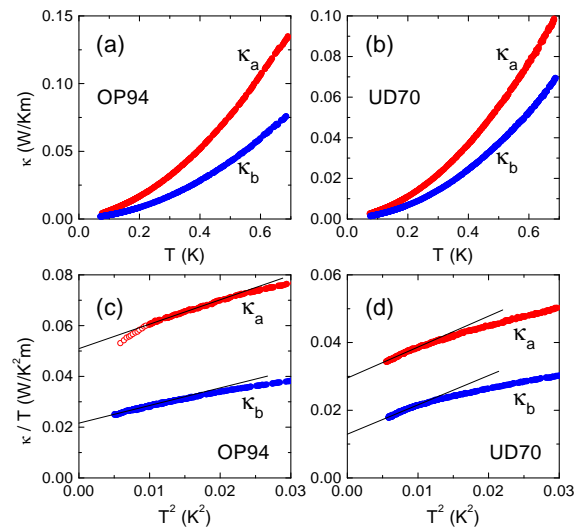} 
\caption{(color online) (a),(b) $T$ dependences of $a$- and 
$b$-axis thermal conductivities for optimally-doped and underdoped 
Bi2212. (c),(d) $\kappa/T$ vs $T^2$ plots of the low-$T$ data; 
the solid lines are linear fits to extract $\kappa_0/T$.} 
\end{figure} 

Figure 4 shows another novel property observed in Bi2212; namely, 
the QPs show very different abilities to conduct heat along the 
$a$ and $b$ axes that correspond to the two orthogonal nodal 
directions. Indeed, the data in Fig. 4 demonstrate that in the $T 
\rightarrow 0$ limit the anisotropy in $\kappa_0/T$ is more than 
a factor of 2. If the universal scenario holds here and Eq. 
(\ref{k0/T}) is valid, the in-plane anisotropy would mean that 
the SC gap is not the standard $d_{x^2-y^2}$, a possibility 
discussed in Ref. \cite{Ando3}; however, given the large 
anisotropy for $T \rightarrow 0$ and the problems in the universal 
scenario, it seems more natural to consider that the source of the 
anisotropy is not an asymmetry in the SC gap but is the breakdown 
of the universal scenario. To corroborate this interpretation, 
the $\kappa_0/T$ values for both $a$ and $b$ axes measured here 
in our optimally-doped Bi2212 are much larger than the values 
reported previously \cite{Nakamae1, Chiao2}; such a dependence of 
$\kappa_0/T$ on the sample source suggests that the details of 
the electronic inhomogeneities matter in Bi2212. 

As for the possible source of the in-plane anisotropy in Bi2212, 
it is useful to notice that the latest ``gap map" data obtained 
by the scanning tunneling microscope experiments by McElroy {\it 
et al.} \cite{McElroy} present clear anisotropy in the electronic 
inhomogeneity; in Fig. 2B of Ref. \cite{McElroy}, the variation 
of the gap magnitude is rather smooth along the $a$ axis, while 
it is much steeper along the $b$ axis. If a stronger electronic 
inhomogeneity causes a larger tendency for QP localization, as 
the calculations based on the Bogoliubov-de Gennes equation 
suggest \cite{Hussey_Review, Atkinson}, one would expect 
$\kappa_b$ to become smaller than $\kappa_a$ in the situation as 
indicated in Ref. \cite{McElroy}. Hence, the unusual in-plane 
anisotropy of $\kappa_0/T$ in Bi2212 is likely to be related to 
the anisotropic electronic inhomogeneity, that is indirectly 
caused by the supermodulation structure peculiar to this system. 

A clear message one can take is that the QP thermal conductivity 
is unexpectedly sensitive to the electronic inhomogeneity, 
whatever the nature of the inhomogeneity is; it can be the local 
suppression of the SC gap by Zn impurities \cite{Pan}, or it can 
be the gap inhomogeneity produced by dopant atoms \cite{McElroy}. 
The qualitative change in the Zn-substitution dependence near 
optimum doping suggests that a strong tendency to form an 
electronic inhomogeneity is present only in the under- to 
optimally-doped regime; in this regard, it is useful to note that 
the dynamics of photo-induced QPs in our Bi2212 crystals shows a 
sharp change at optimum doping, and a more BCS-like behavior is 
found in the overdoped regime \cite{Gedik}. In any case, for a 
better understanding of the QP properties in cuprates, one should 
definitely take into account the QP localization due to 
electronic inhomogeneities, and the present work demonstrates a 
crucial role of the nanoscale inhomogeneity in the low-energy 
physics of high-$T_c$ cuprates. 

\begin{acknowledgments} 

We greatly thank K. Behnia, P. J. Hirschfeld, N. E. Hussey and J. 
Takeya for helpful discussions. This work was supported by the 
Grant-in-Aid for Science provided by the Japan Society for the 
Promotion of Science. 

\end{acknowledgments}


\begin{thebibliography}{}
\bibitem{note_slope} 
While the slope of the linear fit is usually determined by the cross
section of the sample due to the boundary scattering of phonons, the
large slopes in our Zn-free YBCO are indicative of the $T^3$
contribution of the thermally-excited QPs in very clean samples
\cite{Hill}, which testifies the ``ultraclean" nature of our YBCO
crystals. This additional $T^3$ term does not affect the determination
of $\kappa_0/T$.
\end{thebibliography}
\end{document}